\documentclass[final]{aipproc}
\layoutstyle{8x11double}
\usepackage{graphicx,amsmath}
\begin{document}
\title{Global NLO Analysis of \\ Nuclear Parton Distribution Functions}


\author{M. Hirai}{
  address={Department of Physics, Juntendo University, 
           Inba, Chiba, 270-1695, Japan}}
   
\author{S. Kumano}{
  address={Institute of Particle and Nuclear Studies, 
          High Energy Accelerator Research Organization (KEK) \\
          1-1, Ooho, Tsukuba, Ibaraki, 305-0801, Japan},
  altaddress={Department of Particle and Nuclear Studies,
           Graduate University for Advanced Studies \\
           1-1, Ooho, Tsukuba, Ibaraki, 305-0801, Japan}}

\author{T.-H. Nagai}{
  address={Department of Particle and Nuclear Studies,
           Graduate University for Advanced Studies \\
           1-1, Ooho, Tsukuba, Ibaraki, 305-0801, Japan}}
   
\begin{abstract}
Nuclear parton distribution functions (NPDFs) are determined by a global
analysis of experimental measurements on structure-function ratios
$F_2^A/F_2^{A'}$ and Drell-Yan cross section ratios
$\sigma_{DY}^A/\sigma_{DY}^{A'}$, and their uncertainties are
estimated by the Hessian method. The NPDFs are obtained in
both leading order (LO) and next-to-leading order (NLO) of $\alpha_s$. 
As a result, valence-quark distributions are relatively
well determined, whereas antiquark distributions at $x>0.2$ and
gluon distributions in the whole $x$ region have large uncertainties.
The NLO uncertainties are slightly smaller than the LO ones;
however, such a NLO improvement is not as significant as the nucleonic case.
\end{abstract}

\maketitle

\section{Introduction}
From measurements on structure-function ratios
$F_2^A/F_2^{D}$, we found that nuclear parton distribution
functions (NPDFs) are not equal to corresponding nucleonic PDFs.
Although there are many analyses on the PDFs in the nucleon,
the NPDFs have not been investigated extensively.
However, such studies are becoming more and more important
in recent years in order to precisely understand measurements
of heavy-ion reactions at RHIC and LHC.
It should lead to a better understanding on properties of
quark-hadron matters.

Such investigations are also valuable for applications
to neutrino reactions. For an accurate determination of
neutrino oscillation, nuclear corrections in the oxygen
nucleus are important. Although the oscillation experiments
are done in a relatively low-energy region, gross properties
of cross sections could be described by using quark-hadron
duality. In future, nuclear structure functions will be investigated
by the MINER$\nu$A project and at neutrino factories.

The unpolarized PDFs in the nucleon have been investigated
for a long time, and they are relatively well determined.
Uncertainties of the determined PDFs were also estimated recently,
and such studies were extended to the polarized PDFs \cite{aac0306}
and fragmentation functions \cite{hkns07}.
On the other hand, we have been investigating parametrization
of the NPDFs and their uncertainties in a similar technique
\cite{hkm01,hkn04} by analyzing experimental data on nuclear
structure-function ratios $F_2^A/F_2^{A'}$ and Drell-Yan cross section
rations $\sigma_{DY}^A/\sigma_{DY}^{A'}$. However, the uncertainty
estimation was limited to the leading order (LO) in the previous
analysis \cite{hkn04}. It is the purpose of this work to
show the NPDFs and their uncertainties in both LO and
next-to-leading order (NLO) and to discuss NLO improvements
\cite{hkn07}.

\section{Analysis method}
First, the functional form of our NPDFs is introduced.
Since nuclear modifications are generally within the 10\%$-$30\% range
for medium and large size nuclei, it is easier to investigate
the modifications than the absolute NPDFs. We define the NPDFs
at the initial $Q^2$ scale ($\equiv Q^2_0$) as
\begin{eqnarray}
f_i^A(x,Q_0^2)=w_i(x,A,Z)f_i(x,Q_0^2),
\end{eqnarray}
where $f_i^A(x,Q_0^2)$ is the parton distribution with
type $i$ ($=u_v$, $d_v$, $\bar u$, $\bar d$, $s$, $g$)
in a nucleus, $f_i(x,Q_0^2)$ is the corresponding
parton distribution in the nucleon, $A$ is mass number,
and $Z$ is the atomic number.
The variable $Q^2$ is given $Q^2=-q^2$ with the virtual photon
momentum $q$ in lepton scattering, and the Bjorken variable $x$
is defined by $x =Q^2/(2M\nu)$ with the energy transfer $\nu$
and the nucleon mass $M$. We call $w_i(x,A,Z)$ a weight function,
which indicates a nuclear modification for the type-$i$ distribution.
The weight functions are expressed by 
\begin{eqnarray}
  w_i(x,A,Z)=1+\Bigl( 1-\frac{1}{A^\alpha} \Bigr )
  \frac{a_i+b_ix +c_i x^2 +d_i x^3}{(1-x)^{\beta_i}},
\end{eqnarray}
where $\alpha$, $a_i$, $ b_i$, $c_i$, $d_i$, and $\beta_i$ are
parameters to be determined by a $\chi^2$ analysis.
Here, the valence up- and down-quark parameters are
the same except for $a_{u_v}$ and $a_{d_v}$.
Since there is no data to find flavor dependence in antiquark
modifications, the weight functions of $\bar{u}$, $\bar{d}$,
and $\bar{s}$ are assumed to be the same at $Q_0^2$.
We impose three conditions, baryon-number, charge, and momentum
conservations, so that three parameters are fixed.

The initial scale of the NPDFs is chosen $Q_0^2=1\ {\rm GeV^2}$.
They are evolved to experimental $Q^2$ points.
From the threshold at $Q^2=m_c^2$, charm-quark distributions
appear due to the $Q^2$ evolution.
Using these NPDFs, we calculate $F_2^A/F_2^{A'}$
and Drell-Yan ratios $\sigma_{DY}^A/\sigma_{DY}^{A'}$.
The parameters are determined so as to minimize the total $\chi^2$
\begin{eqnarray}
\chi^2=\sum_j \frac{(R_j^{data}-R_j^{theo})^2}{(\sigma_j^{data})^2},
\end{eqnarray} 
where $R_j$ indicates $F_2^A/F_2^{A'}$ and $\sigma_{DY}^A/\sigma_{DY}^{A'}$.
The uncertainties of the determined NPDF are estimated by the Hessian method:
\begin{eqnarray}
[\delta f^A (x)]^2 =\Delta \chi^2 \sum_{i,j} 
\frac{\partial f^A (x,\hat{\xi})}{\partial \xi_i} H_{ij}^{-1} 
\frac{\partial f^A(x,\hat{\xi})}{\partial \xi_j} ,
\end{eqnarray}
where $H_{i j}$ is the Hessian matrix, $\xi_i$ is a parameter, 
and $\hat\xi$ indicates the optimum parameter set.
The $\Delta \chi^2 $ value is taken as 13.7 so that the confidence
level $P$ becomes the one-$\sigma$-error range ($P=0.6826$)
for thirteen parameters by assuming the normal distribution
in the multi-parameter space.

\section{Results}
We used data with $Q^2 \ge 1\,{\rm GeV^2}$. They consist of 
290, 606, 293, and 52 data points for $F_2^D/F_2^p$, $F_2^A/F_2^D$,
$F_2^A/F_2^{A'}$ ($A' \ne D$), and $\sigma_{DY}^A/\sigma_{DY}^{A'}$,
respectively. For the total 1241 data, we obtained the minimum $\chi^2$
values, $\chi_{min}^2/$d.o.f.=1.35 and 1.21 for the LO and NLO.

\begin{figure}[t]
        \includegraphics*[width=0.7\hsize]{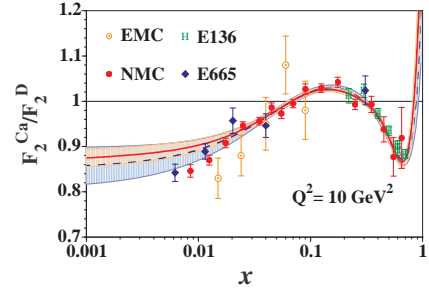} 
\caption{Experimental data of $F_2^{Ca}/F_2^D$ are compared
with theoretical ratios, which are
calculated at $Q^2=10\ {\rm GeV^2}$. 
The dashed and solid curves indicate LO and NLO results,
and the uncertainties are shown by the shaded bands.}
\label{fig:F2CaD} 
\end{figure}
\begin{figure}[t!]
        \includegraphics*[width=0.7\hsize]{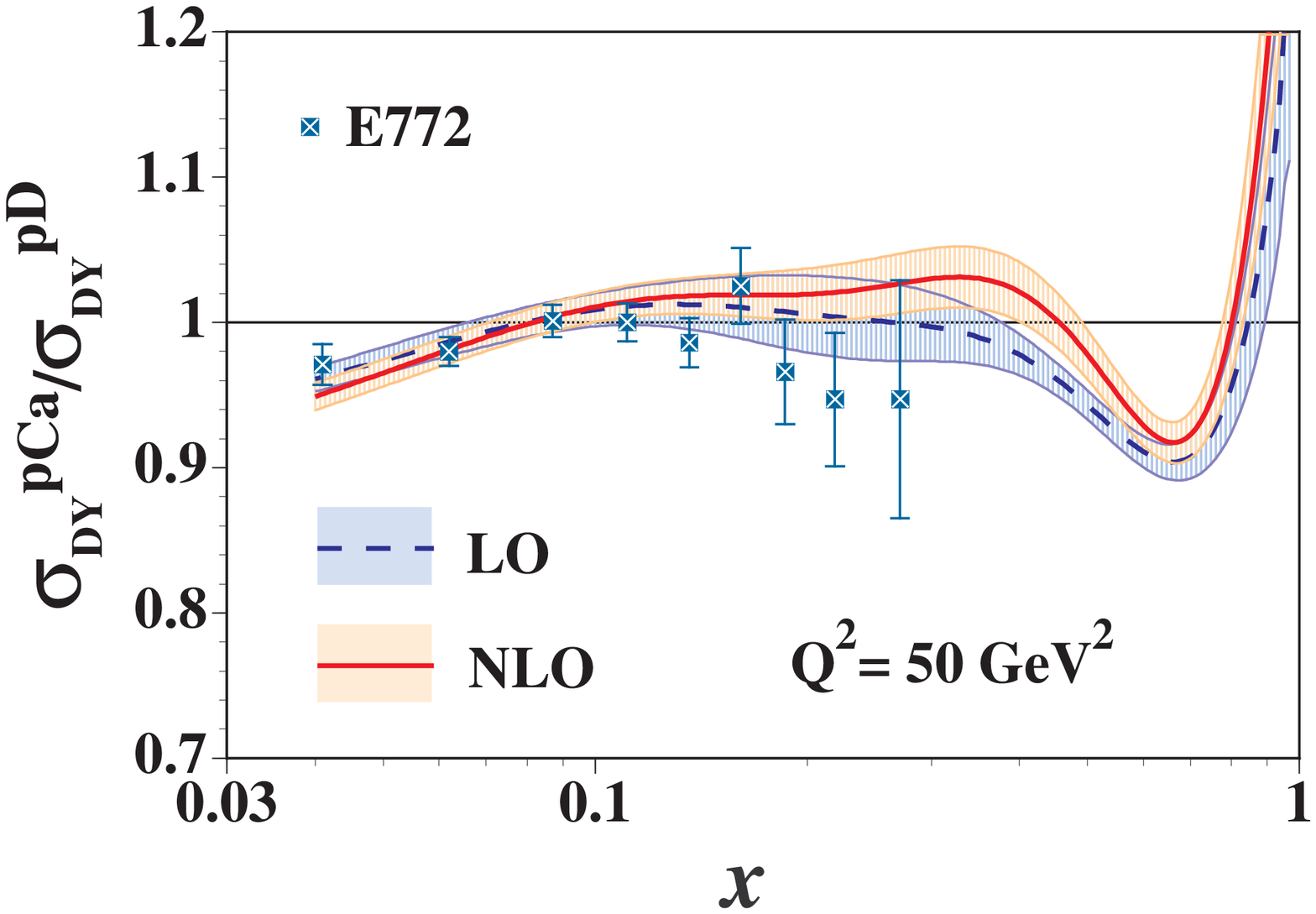} 
\caption{Experimental data of $\sigma_{DY}^{p Ca}/\sigma_{DY}^{pD}$
are compared with the theoretical ratios calculated at
$Q^2=50\ {\rm GeV^2}$. The dashed and solid curves indicate
LO and NLO results, and the uncertainties are shown
by the shaded bands.}
\label{fig:DYCa} 
\end{figure}

As examples, we show actual data with the theoretical LO and NLO
ratios for the calcium nucleus together with their uncertainties
in Figs. \ref{fig:F2CaD} and \ref{fig:DYCa}. The theoretical
curves and their uncertainties are calculated at $Q^2=10\ {\rm GeV^2}$
and $Q^2=50\ {\rm GeV^2}$ for the $F_2$ and the Drell-Yan, respectively;
however, the experimental data were taken at various $Q^2$ values.
There are discrepancies from the experimental data at $x<0.01$
in Fig. \ref{fig:F2CaD}, but they should be attributed to
the $Q^2$ difference \cite{hkn07}.
The figures indicate good agreement between the data and
the theoretical curves, and most data are within the uncertainties.
As expected, the NLO uncertainties are smaller than the LO ones,
especially at small $x$, but they are similar in the region, $x>0.02$. 
As for Drell-Yan ratios in the region $x>0.04$, the NLO effects
on the uncertainties are not obvious.  

\begin{figure}[t]
        \includegraphics*[width=0.85\hsize]{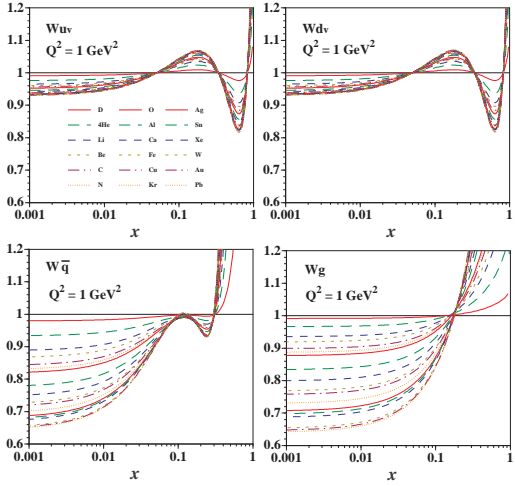} 
\caption{Nuclear modifications $w_i$ ($i=u_v$, $d_v$, $\bar{q}$,
and $g$) are shown in the NLO for all the analyzed nuclei
and $^{16}\rm O$ at $Q^2 = 1\ {\rm GeV^2}$.}
\label{fig:all-npdfs} 
\end{figure}

In order to illustrate the nuclear dependence, we show 
the weight functions for all the analyzed nuclei and $^{16}\rm O$
at $Q^2 = 1\ {\rm GeV^2}$ in Fig. \ref{fig:all-npdfs}.
The NPDFs in the oxygen nucleus are shown because they are
important for neutrino-oscillation studies.
As the mass number becomes larger in the order of D, $\rm ^{4}He$,
Li, $\cdots$, and Pb, the curves deviate from the line of unity. 

We provide a code for calculating the NPDFs and their uncertainties
at our web site \cite{npdfweb}. By supplying the kinematical conditions,
$x$ and $Q^2$, and a nuclear species, one can obtain the NPDFs
($u^A$, $d^A$, $s^A$, $\bar u^A$, $\bar d^A$, $s^A$, $c^A$, and $g^A$)
numerically. The technical details on its usage
are explained in Refs. \cite{hkn04,hkn07} and within the subroutine. 

\begin{figure}[b]
        \includegraphics*[width=0.85\hsize]{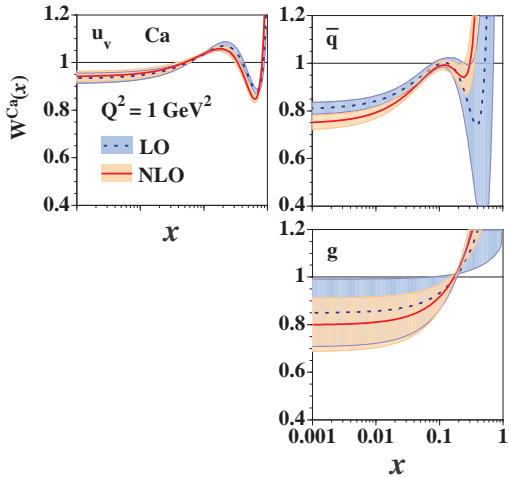} 
\caption{Nuclear modifications of the PDFs and their uncertainties
         are shown for the calcium nucleus at $Q^2 = 1\ {\rm GeV^2}$.
         The dashed  and solid curves indicate LO and NLO results,
         and their uncertainties are shown by the shaded bands.}
\label{fig:w-Ca} 
\end{figure}

Nuclear modifications and their uncertainties are shown for
the calcium nucleus in Fig. \ref{fig:w-Ca}. The valence-quark
distributions are well determined in the medium and large $x$
regions because the $F_2$ structure functions are dominated by 
the valence-quark distributions and because the $F_2$ ratios are
accurately measured. At small $x$, the valence-quark modifications
have also small uncertainties because of the baryon-number and
charge conservations.  

The antiquark distributions are well determined at $x<0.2$ due to 
the $F_2$ and Drell-Yan data. However, they have large uncertainties
at $x > 0.2$ because there is no Drell-Yan data which constrains
the antiquark modifications. 
Future Drell-Yan measurements at large $x$ should improve the situation of
the antiquark distributions. There are J-PARC, Fermilab-E906, and 
GSI-FAIR projects, in which Drell-Yan processes will be investigated.

We found that the uncertainty bands for the gluon are very large,
which indicates that the gluon modifications cannot be well determined
in the whole $x$ region. The gluon distributions contribute
to the $F_2$ and Drell-Yan ratios as higher-order effects.
Therefore, the gluon distributions cannot be accurately determined
especially in the LO analysis. Some improvements are expected in
the NLO analysis. In fact, Fig. \ref{fig:w-Ca} indicates that
the NLO uncertainty band for the gluon becomes smaller than
the LO one. However, it is not as clear as the improvements
in the polarized PDFs \cite{aac0306} and the fragmentation functions
\cite{hkns07}. The gluon distribution in the nucleon has been determined
mainly by using $Q^2$ dependence of the structure function $F_2$.
In the nuclear case, the $Q^2$ dependencies of the ratios
$F_2^A/F_2^{A'}$ are not accurately measured. It leads to the large
uncertainty bands in the gluon modifications even in the NLO analysis.
We hope that accurate data will be provided at future electron
facilities such as eRHIC and eLIC.

Nuclear modifications in the deuteron are also investigated in
our recent analysis \cite{hkn07}. Since the deuteron data are
used for determining the ``nucleonic'' PDFs after
some nuclear corrections, current PDFs in the nucleon could
partially contain nuclear effects. Proper nuclear corrections
should be taken into account in the PDF analysis of the nucleon
to exclude such effects.

\section{Summary}
By the global analyses of the data on the nuclear $F_2$ and Drell-Yan 
ratios, the NPDFs have been determined in both LO and NLO. Although
the valence-quark distributions and antiquark ones at $x<0.2$ are well
determined, the uncertainty bands are very large in the antiquark
modifications at $x > 0.2 $ and the gluon ones at whole $x$
even in the NLO analysis.
We need future experimental efforts for determining all
the nuclear modifications. Our NPDFs and their uncertainties
can be calculated by the code supplied in Ref. \cite{npdfweb}.



\end{document}